\magnification=\magstep1
\hfuzz= 6pt
\baselineskip=16pt
$ $
\vskip 1in
\centerline{{\bf The capacity of the noisy quantum
channel}\footnote{$^\dagger$}{This work supported
in part by
grant \# N00014-95-1-0975 from the Office of Naval
Research.}}

\bigskip
\centerline{Seth Lloyd}

\centerline{D'Arbeloff Laboratory for Information
Systems and Technology}

\centerline{Department of Mechanical Engineering}

\centerline{Massachusetts Institute of Technology}

\centerline{MIT 3-339, Cambridge, Mass. 02139}

\centerline{slloyd@mit.edu}

\bigskip
\noindent {\it Abstract:} An upper limit is given to the
amount of quantum information that can be transmitted
reliably down a noisy, decoherent quantum channel.  A class of
quantum error-correcting codes is presented that allow
the information transmitted to attain this limit.  The
result is the quantum analog of Shannon's bound and code
for the noisy classical channel.

\vskip 1cm

The `quantum' in quantum mechanics means `how much' ---
in quantum mechanics,
classically continuous variables such as energy,
angular momentum and charge come in discrete units
called quanta.
This discrete character of
quantum-mechanical systems such as photons, atoms, and
spins allows them to register ordinary digital information.
A left-circularly polarized photon can encode a 0, for
example, while a right-circularly polarized photon can
encode a 1.  Quantum systems can also
register information in ways that classical digital systems
cannot: a transversely polarized photon is in a quantum
superposition of left and right polarization, and in
some sense encodes both 0 and 1 at the same time.  Even
more surprising from the classical perspective are
so-called entangled states, in which two or more quantum
systems are in superpositions of correlated states, so
that two photons can encode, for example, 00 and 11 at
once.  Such entangled states behave in ways that
apparently violate classical intuitions about locality
and causality (without, of course, actually violating
physical laws).

Information stored on quantum systems that can exist in
superpositions and entangled states is called quantum
information.  The unit of quantum information is the
quantum bit, or qubit (pronounced `Q-bit'),$^1$ the
amount of quantum
information that can be registered on a single two-state
variable such as a photon's polarization or a neutron's
spin.  This paper puts fundamental limits on the amount
of quantum information that can be transmitted
reliably along a
noisy communication channel such as an optical fiber.
Theorems are presented that limit the rate at
which arbitrary superpositions of qubits can be sent
down a channel with given noise characteristics, and
encoding schemes are presented that attain that limit.

It is important to compare the results presented
here---the use of a quantum channel to transmit quantum
information---with schemes that use quantum channels to
transmit classical information, as in Caves and
Drummond's comprehensive review of quantum limits on
bosonic communication rates.$^2$  The limit to the rate
at which arbitrary sequences of ordinary classical bits,
suitably encoded as quantum states,
can be transmitted down a quantum channel such as an optical
fiber is given by Holevo's theorem.  In
contrast, the results presented here limit the rate at
which arbitrary {\it superpositions} of sequences of quantum
bits can be
sent reliably down a noisy, decoherent quantum channel.
As such, the theorems presented in this paper are
complementary to the results of Schumacher$^{1,3}$ and
Josza$^3$ on the noiseless quantum channel.
Any channel that can transmit quantum information can be
used to transmit classical information as well.  It is
possible, however, for a channel to be able to transmit
classical information without being able to transmit
quantum information: examples of such completely
decoherent channels will be discussed below.

The difference between quantum and classical information
does not arise from a fundamental physical distinction
between the systems that register, process, and
transmit that information.  As just noted, quantum
channels can be used to transmit classical information.
And after all, `classical' information-registering
systems such as capacitors and neurons are at
bottom quantum-mechanical.  The difference arises from
the conditions under which such systems operate.  When
properly isolated from their environment, photons
and atoms can exist in superpositions and entangled
states for long periods of time, with experimentally
measurable results.  Capacitors and neurons, in
contrast, interact strongly with a thermal environment,
which prevents them from exhibiting coherent quantum
effects.  As a result, quantum information can be used to
perform tasks that classical information cannot.

A full theory of quantum information and its properties
does not yet exist.  However, the ability to
transmit and process quantum information reliably
provides the solution to problems to which no classical
solution is known: if entangled quantum bits can be
transmitted and received, quantum cryptographic
techniques can be used to create provably secure shared
keywords for unbreakable codes;$^{4}$ while the ability to
process quantum information allows quantum computers
efficiently to factorize large numbers and to simulate
local quantum systems.$^{5}$

For quantum information to prove useful, it must be
transmitted and processed {\it reliably}.  Quantum
superpositions and entangled states tend to be easily
disrupted by noise and by interactions with their
environment, a process called decoherence.$^{6-7}$  Until
recently, decoherence and noise seemed insurmountable
obstacles to reliable quantum information transmission
and processing.  However, in 1995, Shor exhibited the
first quantum error-correcting routine.$^{8}$  since then,
several such routines have been proposed$^{9-13}$  All
of these routines have the feature, common to many
classical error-correcting codes as well, that the
rate of transmission of quantum information goes to
zero as the reliability of transmission goes to one. This
paper shows that arbitrarily complicated quantum states
can in principle be encoded,
subjected to high levels of noise and decoherence, then
decoded to give a state arbitrarily close to the
original state, all with a finite rate of transmission
of quantum information.  The paper states and outlines
the proof
of theorems that put on upper bound to the capacity of
noisy, decoherent quantum channels to transmit quantum
information reliably, and exhibits a class of quantum
codes that attain that bound.

\bigskip\noindent {\bf 1. Quantum Sources}

A quantum channel has a {\it source} that emits systems
in quantum
states, (the {\it signal}) to the channel, and a
receiver that receives the noisy, decohered signal
emitted by the channel.  For example, the source could
be a highly attenuated laser that emits individual
monochromatic photons, the channel could be an optical
fiber, and the receiver could be a photocell.  Or the
source could be a set of ions
in an ion-trap quantum computer$^{14}$ that have been
prepared by a sequence of laser pulses in an entangled
state, the channel could be the ion trap in which the
ions evolve over time, and the receiver could be a
microscope to read out the states of the ions via
laser-induced fluorescence.  This second example
indicates that a quantum channel can transmit quantum
information from one time to another as well as from one
place to another.  As Shannon emphasized, a computer
memory is a communications channel.

A more complete picture of a quantum channel is
as follows (Figure 1): the input signal is some unknown
quantum state; the input is fed into an encoder
that transforms it into a redundant form;
the encoded signal is sent down the channel,
subjected to noise and decoherence; the noisy,
decohered signal is then fed into a decoder that
attempts to restore the original signal.
Quantum encoding and decoding
requires the ability to manipulate quantum states in a
systematic fashion, for example, by using Kimble's$^{15}$
photonic quantum logic gates or
Wineland's realization$^{14}$ of the
ion-trap quantum computer
proposed by Cirac and Zoller.$^{16}$
{}From a practical point of view, such decoding and
encoding may prove the most difficult part of reliable
quantum information transmission and processing.  This
paper will simply exhibit coding and decoding
schemes that attain the channel capacity:
it will not address how such schemes can be
carried out in practice.

In order to demonstrate the quantum analog of Shannon's
noisy coding theorem,$^{17}$ it's helpful to set up a quantum
formalism that corresponds closely to the classical
picture of a noisy channel.  Quantum systems and
quantum signals are described by states
$|\psi\rangle$ in a Hilbert space ${\cal H}$, or more
generally, by density matrices $\rho\in
{\cal H}^*\otimes{\cal H}.$
A quantum ensemble ${\cal
E}= \{(|\psi_i\rangle, p_i)\}$
is a set of quantum states
$|\psi_i\rangle$ belonging to the same Hilbert space
${\cal H}$, together with their probabilities $p_i$.
The expectation value of a measurement on the ensemble
corresponding to a Hermitian operator $M$ is $\langle M
\rangle_{\cal E} = \sum_i p_i \langle \psi_i
|M|\psi_i\rangle$ $= {\rm tr} M\rho_{\cal E}$, where
$\rho_{\cal E}=\sum_i p_i |\psi_i\rangle\langle\psi_i|$
is the density matrix corresponding to
the ensemble.  The states $|\psi_i\rangle$ need be
neither orthonormal nor normalized, as long as $\sum_i
p_i \langle \psi_i|\psi_i\rangle = {\rm tr} \rho_{\cal
E} = 1$.  That is, a quantum ensemble is just the
quantum analog of a classical ensemble, where care has
been taken to take into account the inherently
statistical nature of quantum mechanics.

Two ensembles that have the same density matrix are
statistically indistinguishable: no set of measurements
can distinguish whether a sequence of states is drawn
from one ensemble rather than the other.  An example of
statistically indistinguishable ensembles is ${\cal E}_1
= \{(|\uparrow\rangle, 1/2) ~,~
(|\downarrow\rangle, 1/2) \}$, and
$${\cal E}_2 = \{(|\uparrow\rangle, 1/3)~,~
(1/2|\uparrow\rangle + \sqrt{3}/2|\downarrow\rangle,
1/3)~,~ (1/2|\uparrow\rangle - \sqrt{3}/2|\downarrow\rangle,
1/3)\}\quad,$$
\noindent both with density matrices $\rho= 1/2
|\uparrow\rangle\langle \uparrow| + 1/2
|\downarrow\rangle\langle\downarrow|$.  Note that an
ensemble over a finite dimensional Hilbert space can
contain an infinite number of states, e.g., ${\cal E} =
\{ (e^{i\phi}|\uparrow\rangle, p(\phi) = 1/2\pi )\}$, in
which case each state is paired with a continuous
probability density, $p(\phi)$, and $\rho=
\int_0^{2\pi} (1/2\pi)e^{i\phi}|\uparrow\rangle \langle
\uparrow| e^{-i\phi} d\phi = |\uparrow\rangle \langle
\uparrow|$.  Because of the inherently statistical
nature of quantum mechanics, different quantum ensembles
can be statistically indistinguishable,
while two classical
ensembles are statistically indistinguishable if and
only if they are identical.

A particularly interesting type of continuous quantum
ensemble is the uniform ensemble over a Hilbert space ${\cal
H}$, ${\cal E}_{\cal H}= \{(|\phi\rangle\in {\cal H},
p_\phi= 1/{\rm vol}{\cal H})\}$, where ${\rm
vol}{\cal H}$ is the volume of the unit sphere in
${\cal H}$.  This ensemble contains every possible
state and superposition of states in ${\cal H}$, all with
equal probabilities.  The corresponding density matrix
is $\rho_{\cal H} = (1/d)
\sum_{i=1}^d |\phi_i\rangle \langle
\phi_i|$, where $d$ is the dimension of
${\cal H}$ and $\{|\phi_i\rangle\}$ is an orthonormal
basis for ${\cal H}$.  If we wish to transmit
arbitrary superpositions of states down
quantum channels, the sources of interest are of the
form ${\cal E}_{\cal H}$ for some ${\cal H}$.

Like Shannon, we will restrict our attention to
stationary, ergodic sources.$^{17}$  A stationary
source is one for which the probabilities for emitting
states doesn't change over time; an ergodic source is
one in which each sub-sequence of states appears in
longer sequences with a frequency equal to its
probability.
(These assumptions are made for convenience
of analysis only: in fact, the inherently statistical
nature of quantum mechanics makes them less necessary in
the quantum than in the classical case, and the results
derived can be generalized to non-stationary,
non-ergodic sources.)
We define a stationary, ergodic
ensemble over $N$ time steps as one whose density
matrix is the tensor product of $N$ times its density
matrix over a single time step: $\rho^N = \rho\otimes
\rho\otimes \ldots \otimes\rho$.

There are many different quantum ensembles with
density matrix $\rho\otimes\ldots \otimes\rho$.  But
as noted by Schumacher$^{1,3}$ and Josza $^3$,
there is one ensemble in particular that effectively
contains all such ensembles.
Let $\rho=\sum_i p_i |\phi_i\rangle\langle \phi_i|$,
where the ${\phi_i}$ are orthonormal.
Consider the subspace $\tilde{\cal H}^N$ spanned by
the `high-probability' product states $|\phi_{i_1}\rangle
\ldots |\phi_{i_N}\rangle$, where each $|\phi_i\rangle$
occurs in the product $\approx p_iN$ times.  These
states are the analog of high-probability sequences
of symbols for a classical source.
The following theorem then follows as an immediate
corollary to the noiseless quantum channel source
theorem of Schumacher$^{1,3}$ and Josza$^3$:

\bigskip\noindent{\bf Theorem 1.} (Quantum source
theorem.)  Let $|\psi\rangle$ be selected
from {\it any} ensemble with
density matrix $\rho\otimes\ldots\otimes\rho$.  Then
as $N\rightarrow\infty$, $|\psi\rangle$ is to be found
in the high-probability subspace
$\tilde{\cal H}_N$ with probability 1.
$\tilde{\cal H}_N$ is a {\it minimal}
subspace with this property, in the sense that any other
such subspace contains $\tilde{\cal H}_N$.

\bigskip
That is, as $N\rightarrow\infty$, the
ensemble ${\cal E}_{\tilde{\cal H}^N}$ contains with
probability one the members of any ensemble with
density matrix $\rho\otimes\ldots\otimes\rho$.
A more precise statement of theorem 1 is that
as $N\rightarrow\infty$,
$\sum_{|\psi\rangle} p_{|\psi\rangle}
\langle \psi| P_{\tilde{\cal H}_N} |\psi\rangle
\rightarrow 1$, where
$P_{\tilde{\cal H}^N}$ is the projection operator onto
$\tilde{\cal H}^N$.
By Shannon's source theorem, the dimension
of ${\cal E}_{\tilde{\cal H}^N}$ is approximately
$e^{NS}$ where $S=-{\rm tr}\rho {\rm ln} \rho$.
As with Shannon's theorems for classical sources,
which simplify the analysis of the classical noisy
channel by focusing on high-probability inputs,
and as with the use of high-probability subspaces
in the noiseless quantum channel theorem
in references (1) and (3),
the quantum source theorem simplifies
the analysis of the noisy quantum channel
by focusing on a particular subspace of
inputs.  A coding scheme that works for any ensemble
with density matrix $\rho\otimes\ldots\otimes\rho$
works for the states in the high probability subspace.
Conversely, a coding scheme that works for the
high-probability subspace works for any of the
ensembles that it contains.  Accordingly, from this
point on, quantum sources will be taken to be
ensembles over high-probability subspaces unless
otherwise stated.

\bigskip\noindent{\bf 2. The Quantum Channel}

A quantum communications channel takes quantum
information as input and produces quantum information
as output.  An optical fiber is an example of a quantum
channel: a photon in some quantum state goes in, suffers
noise and distortion in passing through the fiber, and if
it is not absorbed and does not tunnel out, emerges in a
transformed quantum state.  In the normal formulation of
quantum mechanics, the ingoing system that carries quantum
information is described by a density matrix $\rho_{\rm
in}$, and the outgoing system is described by a density
matrix $\rho_{\rm out}={\cal S}(\rho_{\rm in})$, where
${\cal S}$ is a trace-preserving linear operator called a
super-scattering operator.  For simplicity, the channel
will be assumed to be time-independent and memoryless,
so that it has the same effect on each quantum bit that
goes through.  (The generalization to time-dependent
channels with memory is straightforward.)

An equivalent method of formulating the channel's dynamics
specify its effect on each of an orthonormal basis
$\{|\phi_i\rangle\}$ of input states: the output of the
channel for input $|\phi_i\rangle$ is then given by the
ensemble ${\cal E}_{|\phi_i\rangle}= \{
(|\psi_{j(i)}\rangle, p_{j(i)}) \}$ of output states
into which $|\phi_i\rangle$ can evolve, together with the
probabilities $p_{j(i)}$ that $|\phi_i\rangle$ evolves
into the state $|\psi_{j(i)}\rangle$.  The density matrix
and ensemble pictures of the effect of the channel are
related as follows: ${\cal S}(|\phi_i\rangle\langle
\phi_{i'} |) = \sum_{j(i)} \sqrt{p_{j(i)}p_{j(i')}}
|\psi_{j(i)}\rangle\langle \psi_{j(i')} |$, which for
$i=i'$ gives ${\cal S}(|\phi_i\rangle \langle \phi_i|)
\sum_{j(i)}|\psi_{j(i)}\rangle\langle \psi_{j(i)}|.$

For example, if the channel is noiseless and
distortion-free, then ${\cal S}$ is the identity operator,
and ${\cal E}_{|\phi_i\rangle} = \{(|\phi_i\rangle,
1)\}$.  This channel transmits both classical and
quantum information perfectly.  Another example is the
completely decohering channel, which can be thought of as
the channel that destroys off-diagonal terms in the
density matrix: ${\cal S}(\sum_{ij} \alpha_{ij}
|\phi_i\rangle\langle \phi_j|) = \sum_{i} \alpha_{ii}
|\phi_i\rangle\langle \phi_i|$, or equivalently and
perhaps more intuitively, as the channel that randomizes
the phases of input states: $|\phi_i\rangle
\longrightarrow {\cal E}_{|\phi_i\rangle} = \{
(e^{i\lambda} |\phi_i\rangle, p(\lambda)= 1/2\pi) \}$.
The completely decohering channel highlights the
difference between the use of quantum channels to carry
classical information and their use in carrying quantum
information: it transmits classical information perfectly,
but transmits no quantum information at all --- no
superpositions or entanglements survive transmission.

Most quantum channels are neither noiseless nor completely
decohering.  The next theorem quantifies just how much
quantum information can be sent down a noisy, decohering
channel.  As above, we restrict our attention to
stationary ergodic sources with density matrix $\rho_{\rm
in} = \sum_ip_i |\phi_i\rangle \langle \phi_i|$.  The inputs
to the channel are then described by a density matrix
$\rho^N_{\rm in} = \rho_{\rm in}\otimes \ldots \otimes
\rho_{\rm in}$, and the output is described by a density
matrix $\rho^N_{\rm out} = \rho_{\rm out} \otimes \ldots
\otimes \rho_{\rm out}$, where $\rho_{\rm out} = {\cal S}(
\rho_{\rm in}) = \sum_{i,j(i)} p_ip_{j(i)}
|\psi_{j(i)}\rangle \langle \psi_{j(i)}|$.

As $N\rightarrow\infty$, input states
come from the subspace $\tilde{\cal H}_{\rm in}^N$ with
probability 1, and output states lie in the subspace
$\tilde{\cal H}^N_{\rm out}$ spanned by high-probability
sequences of outputs, $|\psi_{j_1(i_1)}\rangle \ldots
|\psi_{j_N(i_N)}\rangle$, where each $|\psi_{j(i)}\rangle$
appears in the sequence $\approx p_ip_{j(i)} N$ times.
The dimension of $\tilde{\cal H}^N_{\rm out}$ is
$\approx 2^{-N{\rm tr}\rho_{\rm out}{\rm log}_2
\rho_{\rm out}}$.
To gauge the quantity of quantum information sent down
the channel, look at the effect of the channel on a
typical input state $|\alpha_N\rangle = \sum_{i_1\ldots
i_N} \alpha_{i_1\ldots i_N} |\phi_{i_1}\rangle \ldots
|\phi_{i_N}\rangle~\in~ \tilde{\cal H}^N_{\rm in}$,
where the sum is over high-probability input sequences
in which $|\phi_i\rangle$ appears $\approx p_iN$ times.
We have,

\bigskip\noindent{\bf Theorem 2:} (Quantum channel
theorem.)  As
$N\rightarrow\infty$, when $|\alpha_N\rangle$ is input
to the channel, the output lies with probability 1 in a
minimal subspace $\tilde{\cal H}^N_\alpha$ whose
average dimension over $\alpha_N$ is
the minimum of $e^{NS_{\rm out}},
e^{NS_{\bar\alpha}}$, where $S_{\bar\alpha}
= -{\rm tr}\rho_{\bar\alpha}{\rm ln} \rho_{\bar\alpha}$
and $\rho_{\bar\alpha}= \sum_{i,i'} \sqrt{p_ip_{i'}} {\cal
S}(|\phi_i\rangle\langle\phi_{i'}|)\otimes |\phi_i\rangle
\langle \phi_{i'}|.$
\bigskip

The proof of theorem 2 is somewhat involved, but the
form of $\rho_{\bar\alpha}$ can be understood simply.
One of the primary uses of a quantum channel is the
distribution of entangled quantum states for the purpose
of quantum cryptography or teleportation.  Take a
two-variable
entangled state of the form $\sum_i\sqrt{p_i}
|\phi_i\rangle |\phi_i\rangle$.  Like the state
$(1/\sqrt{2})(|0\rangle|0\rangle + |1\rangle
|1\rangle)$ described in the introduction, this state is a
maximally entangled state that registers all the
states $|\phi_i\rangle|\phi_i\rangle$ at once; the
factors of $\sqrt{p_i}$ insure that each of the two
quantum variables taken on its own
is described by a density matrix $\rho_{\rm in}$.  Now
send the first variable down the channel.  The result is
a partially entangled state for the two variables
described by density matrix
$\rho_{\bar\alpha}$.  That is, $S_{\bar\alpha}$ is the
entropy increase when one of two fully entangled
variables is sent down the channel.  A thorough
treatment of the effect of
noisy channels on entangled states can be found
in reference (18).
The effect of the channel on an $N$-variable state
$|\alpha_N\rangle$ can be understood as follows: almost
all input states $|\alpha_N\rangle$ are fully entangled,
with density matrix $\rho_{\rm in}$ describing each
variable on its own.$^{19}$ Sending $n$ of the variables
through the channel then increases the entropy by
$nS_{\bar\alpha}$, which is in turn the logarithm of the
dimension of the minimal subspace that can
encompass the channel's possible outputs.  If
$S_{\bar\alpha} > S_{\rm out}$, then sending all
the variables through completely randomizes the
output as $N\rightarrow\infty$,
and no coherent quantum information survives
the transmission through the channel.

Theorem 2 suggests that the amount of quantum
information transmitted down the channel from a
stationary, ergodic source with density matrix $\rho_{\rm
in}$ be defined as
$I_Q(\rho_{\rm in}) =
-{\rm tr} \rho_{\rm out} {\rm
log}_2 \rho_{\rm out} + {\rm tr}\rho_{\bar\alpha} {\rm
log}_2 \rho_{\bar\alpha}= S_{\rm out}-S_{\bar\alpha}$
if $S_{\rm out}> S_{\bar\alpha}$, $
= 0$ otherwise.
This definition of quantum information
transmitted is the quantum analog of mutual information
between channel inputs and outputs: when pure states are
sent down the channel, $I_Q$ tells how much information
one gets about which pure state $\in \tilde{\cal
H}^N_{\rm in}$ went in by looking at the noisy mixed
state $\in \tilde{\cal H}^N_{\rm out}$ that comes
out.$^{20}$

The full justification of $I_Q$ as the quantum
information transmitted down a quantum channel will be
presented in the next section, in which quantum coding
schemes will be presented that allow the reliable
transmission of quantum information at a rate $I_Q$, and
in which it will be noted that no coding schemes exist
for stationary, ergodic sources that can surpass this
rate.  For the moment, consider three examples of
quantum channels, each with source described by
$\rho_{\rm in}= (1/2)(|0\rangle\langle0| +
|1\rangle\langle 1|)$. (i) In the noiseless quantum
channel, $-{\rm tr} \rho_{\rm out} {\rm log}_2 \rho_{\rm
out} = 1$, $-{\rm tr} \rho_{\bar\alpha} {\rm log}_2
\rho_{\bar\alpha} =0$, and $I_Q = 1$ qubit, reflecting
the fact that each qubit is received as sent.  (ii) In the
completely decohering/dephasing channel,
$-{\rm tr} \rho_{\rm out} {\rm log}_2 \rho_{\rm
out} = 1$, $\rho_{\bar\alpha}= (1/2) (|0\rangle\langle
0|\otimes |0\rangle\langle 0| +  |1\rangle\langle
1| \otimes |1\rangle\langle 1|)$,
$-{\rm tr} \rho_{\bar\alpha} {\rm log}_2
\rho_{\bar\alpha} =1$, and $I_Q = 0$ qubits, so that no
quantum information is sent.  (iii) Consider a partly
dephasing channel in which $|0\rangle\langle 0|
\rightarrow |0\rangle\langle 0|$, $|1\rangle\langle 1|
\rightarrow |1\rangle\langle 1|$, and
$|0\rangle\langle 1|
\rightarrow (1-\epsilon)|0\rangle\langle 1|$,
$|1\rangle\langle 0|
\rightarrow (1-\epsilon)|1\rangle\langle 0|$.  Here,
$$
\rho_{\bar\alpha}= (1/2)(|0\rangle\langle
0|\otimes |0\rangle\langle 0| + |1\rangle\langle
1| \otimes |1\rangle\langle 1|)
+(1-\epsilon)(|1\rangle\langle0|\otimes
|1\rangle\langle0|
+|0\rangle\langle1|\otimes
|0\rangle\langle1|) $$
\noindent and $-{\rm tr}\rho_{\bar\alpha} {\rm log}_2
\rho_{\bar\alpha} = -(1-\epsilon/2) {\rm log}
(1-\epsilon/2) - (\epsilon/2){\rm log}_2(\epsilon/2)$,
giving an $I_Q$ that ranges continuously
from 1 for $\epsilon=0$ (no
decoherence) to 0 for $\epsilon=1$ (complete
decoherence).

\bigskip\noindent{\bf 3. Optimal codes for
the noisy quantum channel
}

Define the capacity of a quantum channel to carry
quantum information to be
$C_Q= {\rm max}_{\rho_{\rm
in}} I_Q(\rho_{\rm in}).$
$C_Q$ is the maximum over all sources
$\rho_{\rm in}$ of the quantum information $I_Q$
transmitted down the channel.  We then have the
following

\bigskip\noindent{\bf Theorem 3.} (Noisy quantum
channel coding
theorem.)  Consider a quantum channel with capacity
$C_Q$.  The output of a stationary, ergodic source with
density matrix $\rho$ can be encoded, sent down the
channel, and decoded with reliability $\rightarrow 1$
as $N\rightarrow \infty$ if and only if $-{\rm tr} \rho
{\rm log}_2 \rho \leq C_Q.$
\bigskip

Like Shannon's noisy coding theorem, theorem 3 comes
with the caveat that it applies to high-probability
sources.$^{21}$
The proof to theorem 3 will be given elsewhere: but
the idea behind the proof, as well as
the theorem's meaning and implications can be understood as
follows.  The noisy, decohering quantum channel has
two effects on the quantum information that it
transmits.  First, like the classical channel, it
adds noise to the signal, flipping qubits and adding
random information.  Second, it decoheres the
signal by randomizing phases and acquiring
information about the quantum information
transmitted.  Decoherence is an effect with no
classical analog: classical signals do not have
phases, and acquiring information about a
classical signal is harmless as long as the signal is not
altered in the process.  In quantum mechanics,
however, acquiring information about the signal means
effectively making a measurement on it, and quantum
measurement unavoidably alters most quantum systems.

The problem of decoherence implies that signal
must be encoded in such a way that any
information the channel gets about the encoded state
reveals nothing about which
state of the source was sent.  Otherwise, the channel can
effectively `measure' the output of the source,
irretrievably disturbing it in the process.  As noted
by Shor$^{8}$, this may be accomplished by encoding
the signal as an entangled state.  In fact, each
encoded signal must have the same density
matrix $\rho_{\rm in}$ as each other encoded signal
for each qubit
sent down the channel: otherwise the
channel can distinguish between different signals
and decohere them.
If the signals are encoded as entangled states in
this fashion, the channel can decohere the codeword,
but it cannot decohere the original signal.

Suppose someone hands you a quantum system in some
unknown state selected from an ensemble with density
matrix $\rho\otimes\ldots\otimes\rho$, and asks you
to transmit it reliably down a noisy, decoherent
quantum channel.  What do you do?  (If someone hands
you a system in a known quantum state, no
quantum channel is necessary: you can just use
a classical channel to transmit instructions for
recreating the state using a quantum computer.)
The following encoding attains the channel
capacity:  First, identify a source
for the channel
that attains the channel capacity, so that
$I_Q(\rho_{\rm in}) = C_Q$.
Next, encode the state to be transmitted by applying
a transformation that maps
an orthonormal basis for the input
high-probability subspace to a {\it randomly chosen}
set of orthogonal states taken from
the high-probability subspace of the source that
attains the channel capacity.  Such random states
have the desired property that they are fully
entangled, and each qubit in the encoded signal
has density matrix
$\rho_{\rm in}$.$^{19}$  Now send the encoded
signal
down the channel.  Because the states are fully
entangled, the channel cannot get any information
about the original pre-encoded state: all the channel
can do to disrupt the encoded state is add entropy
$S_{\rm out} - C_Q$
per symbol transmitted.  That is, the encoding
protects the original state from decoherence; and
as long as $-{\rm tr} \rho{\rm log}_2 \rho \leq C_Q$
there is enough redundancy in the encoded state to
recreate the original state, just as in the classical
case.  This method works equally well if the initial
state is pure, mixed, or entangled with some other
system.

\bigskip
\noindent{\it Examples:} In the three cases
discussed in the
previous section, the channel capacity is just
$I_Q$, as calculated.  The important fact to note
is that even very high levels of decoherence
($\epsilon \rightarrow 1$) can
be tolerated in principle.  A case of
considerable interest is that in which each qubit
system sent down the channel has a probability
$\eta$ of being decohered and randomized.  In
this case,
$$\rho_{\bar\alpha} = \sum_{ii'=0,1}
\big( (1-\eta)/2
|i\rangle\langle i'|\otimes
|i\rangle \langle i'| + (\eta/4)
|i\rangle\langle i|\otimes
|i'\rangle \langle i'|\big)\quad.$$

\noindent $S_{\bar\alpha}$ can be calculated for
this case and is equal to $-(3\eta/4){\rm log}_2
( \eta/4) - (1-3\eta/4){\rm log}_2( 1- 3\eta/4)$,
which is equal to 1 for $\eta\approx.252$.
The highest rate of errors that can be corrected
by an optimal coding procedure is just above
$1/4$ (see also reference (12)).
This example contrasts with
the classical channel, in which arbitrarily high
levels of noise can be tolerated in principle:
quantum coding can correct for arbitrarily high
levels either of noise, or of decoherence, but not
of both together.

\bigskip\noindent{\bf Discussion}

In practice, even if the channel capacity
is not exceeded, the amount of noise and
decoherence that can be tolerated is limited by
the ability to encode and decode: as
$N\rightarrow\infty$, the error in the transmitted
state goes to zero, but the amount of quantum
information processing that must be done to encode
and decode becomes large.  The encoding and decoding
itself must be performed reliably.

The usefulness of
the classical noisy coding theorem is also
limited by coding difficulties:
in particular, random codes are hard to encode
and decode.  In this respect, however, the
quantum theorem has a considerable advantage.
As Shannon noted, random codes are effective
because the bits that make up the signal have
no apparent order.  In the classical case, this
implies that sequences of bits must appear
random.  In the quantum case, however, as long
as the encoded signal is fully entangled, each
qubit in the signal taken on its own appears to be
completely random.  As a result, the code words
themselves may be highly regular: a simple example
of a set of codewords that are easy to encode
and decode, but
are sufficiently random to attain the channel
capacity are $N$ qubit analogs of the familiar
two-qubit entangled states
$$
(1/\sqrt{2})(|01\rangle - |10\rangle),
(1/\sqrt{2})(|01\rangle + |10\rangle),
(1/\sqrt{2})(|00\rangle - |11\rangle),
(1/\sqrt{2})(|00\rangle + |11\rangle).$$
\noindent In the classical case,
random codes are hard to construct.  In the quantum
case, codes that are sufficiently random to attain
the channel capacity can be constructed by a brief
quantum computation.

In conclusion,
this paper has derived fundamental limits on the
amount of quantum information that can be sent
reliably down a quantum channel, and has
exhibited codes that attain those limits.  In
fact, almost all codes attain
those limits.  As with Shannon's classical noisy
coding theorem, the rate of transmission
of quantum information remains finite as
the probability of error goes to zero.

\vfill\eject
\noindent{\bf Footnotes:}

\noindent(1) B. Schumacher {\it Physical Review A}
{\bf 51}, 2738 (1995).

\noindent(2) C. Caves and P.D. Drummond, {\it
Rev. Mod. Phys.} {\bf 66}, 1994.

\noindent(3) R. Jozsa and B. Schumacher, {\it J.
Mod. Optics} {\bf 41}, 2343 (1994).

\noindent(4) C.H. Bennett, {\it Physics Today}
{\bf 48}, 24, 1995.

\noindent(5) P.W. Shor, in {\it Proc. 35th Annual
Symposium on Foundations of Computer Science,} ed.
S. Goldwasser. (IEEE Computer Society Press
Nov. 1994) pp. 124-134.

\noindent(6) W.H. Zurek, {\it Physics Today} {\bf 44},
36, 1991.

\noindent(7) R. Landauer,
{\it Philos. Trans. R. Soc. London, Ser. A} {\bf
353}, 367, 1995;
and in {\it Proceedings of the Drexel--4
symposium on Quantum Nonintegrability --
Quantum Classical Correspondence}, D.H. Feng
and B.L. Hu, eds. (International Press 1996).

\noindent(8) P.W. Shor, {\it Phys. Rev. A} {\bf 52},
2493, 1995.

\noindent(9) A.R. Calderbank and P.W. Shor, to appear
in {\it Phys. Rev. A.}

\noindent(10) A. Steane, to be published in {\it
Proc. Roy. Soc. Lond.}

\noindent(11) R. Laflamme, C. Miquel, J.P. Paz, and W.H.
Zurek, submitted to {\it Phys. Rev. Lett.}, eprint
quant-ph/9604024, 1996.

\noindent(12) C.H. Bennett, D. Divincenzo, J.A. Smolin,
W.K. Wootters, submitted to {\it Phys. Rev. A}, eprint
quant-ph/9604024, 1996.

\noindent(13) A. Ekert and C. Machiavello, eprint
quant-ph/9602022, 1996.

\noindent(14) C. Monroe, D.M. Meekhof, B.E. King, W.M.
Itano, D.J. Wineland, {\it Phys. Rev. Lett.} {\bf
75}, 4714, 1995.

\noindent(15) Q.A. Turchette, C.J. Hood, W. Lange,
H. Mabuchi, H.J. Kimble, {\it Phys. Rev. Lett.}
{\bf 75}, 4710, 1995.

\noindent(16) J.I. Cirac, P. Zoller, {\it Phys. Rev.
Lett.} {\bf 74}, 4091 (1995).

\noindent(17) C.E. Shannon and W. Weaver, {\it The
Mathematical Theory of Communication}, University
of Illinois Press, 1948.

\noindent(18) B. Schumacher, eprint
quant-ph/9604023, 1996.

\noindent(19) S. Lloyd, `Black Holes, Demons, and
the Loss of Coherence,' Ph.D. Thesis, Rockefeller
University, 1988.

\noindent(20) A similar expression for quantum
channel capacity $C_Q$ has been derived independently
by M.A. Nielsen and B. Schumacher, eprint
quant-ph/9604022, who call the quantity `coherent
information.'

\noindent(21) A method for using
measure-zero codes that surpasses the
limits given by theorems 2 and 3 (the codes do not
obey the conditions of the theorems and so do not
contradict them) has recently been
suggested by P.W. Shor and J.A. Smolin.

\vfill\eject

\noindent{\bf Appendix 0}

\noindent Properties of ensembles of states.
The idea behind the the ensemble picture of quantum
mechanics is to deal
with mixtures and superpositions in the same
formalism.  Accordingly, a primary
purpose of the ensemble picture
is to make an explicit distinction
between quantum states that can interfere with
eachother, and quantum states that can't.  The
ensemble picture is constructed so that
different members of an ensemble cannot interfere
with eachother, while corresponding members of
different ensembles can interfere.  The second
purpose of the ensemble picture is to keep track
explicitly of the normalization of states, so that
high-probability sets of states can be identified
correctly.

As noted
on page 4, a quantum ensemble ${\cal E}_\psi = \{
(|\psi_j\rangle, p_j)\}$  is a set of quantum
states together with their probabilities.  Ensembles
are collections of vectors, and share many properties
of vectors.  For example, if ${\cal E}_\psi$
and ${\cal E}_\phi =\{ (|\phi_i\rangle,
q_j )\}$, we can define
a scalar product ${\cal E}_\psi \cdot {\cal E}_\phi =
\sum_j \sqrt{p_jq_j} \langle \psi_j|\phi_j\rangle$.
If ${\cal E}$ is normalized, then ${\cal
E}\cdot{\cal E} = {\rm tr}\rho_{\cal E} =1$.  (Note
that the rule for obtaining the proper statistics is
to associate a factor of $\sqrt{p_j}$ with each
occurrence of $|\psi_j\rangle$.)  This
vector-like character of ensembles allows the
straightforward characterization of properties of
quantum operators.  For example, the trace-preserving
character of the super-scattering operator (page 7)
can be summarized by the requirement that
${\cal E}_{|\phi_j\rangle} \cdot {\cal E}_{|\phi_j'
\rangle} = \delta_{jj'}$.

A type of ensemble that
will prove useful below is one that is obtained by
superposing corresponding states from two ensembles.
If corresponding states have the same probability,
for example if $p_j=q_j$ for the ensembles ${\cal
E}_\phi$, ${\cal E}_\psi$ above, then the ensemble
of superpositions of $\alpha$ times the states of
${\cal E_\phi}$ plus $\beta$ times the
corresponding states of ${\cal E}_\psi$ is just
$\{(\alpha|\phi_j\rangle + \beta|\psi_j\rangle, p_j
)\}$, with density matrix $\rho$ as above.  In
fact, because we will work with ensembles of
high-probability states, which have
equal probabilities, this is the type of
ensemble that we will have occasion to use below.
If the corresponding states from the different
ensembles do not have the same probabilities, then
we write the ensemble of superposed states as
${\cal E}_{\alpha\phi + \beta\psi} = \{( \alpha |\phi_i
\rangle + \beta |\psi_j\rangle, p_jq_j)\}$ to indicate
the ensemble obtained by superposing $\alpha$ times
the states of ${\cal E}_\phi$ plus $\beta$ times the
corresponding states of ${\cal E}_\psi$,
together with a list $p_jq_j$ of the probabilities
of the individual states in the superposition.
We specify superposition ensembles in this fashion
to keep track explicitly of the normalization of the
individual states in the superposition.
The proper overall normalization
of such ensembles is obtained as above by
associating a factor
of $\sqrt{p_j}$ with each $|\psi_j\rangle$ and a
factor of $\sqrt{q_j}$ with each $|\phi_j\rangle$, so
that
$$\rho_{{\cal E}_{\alpha\phi + \beta\psi}} =
\sum_j \alpha\bar\alpha q_j|\phi_j\rangle \langle
\phi_j|+  \alpha\bar\beta \sqrt{q_jp_j} |\phi_j\rangle
\langle\psi_j| + \beta\bar\alpha \sqrt{p_jq_j}
|\psi_j\rangle\langle \phi_j| + \beta\bar\beta p_j
|\psi_j\rangle\langle \psi_j|.$$
Note that the superposition ensemble has the
same density matrix as the ensemble of unnormalized
states,
$\{(\alpha\sqrt{q_j}|\phi_j\rangle + \beta
\sqrt{p_j}|\psi_j\rangle, 1)\}$.
If we wish to superpose many ensembles,
${\cal E}_i =
\{(|\psi_{j(i)}\rangle, p_{j(i)})\}$,
we will use
$i$ to index the ensembles, and $j$ to index the
different members of each
ensemble: e.g., ${\cal E}_\beta=
\{( \sum_i \beta_i|\psi_{j(i)}\rangle, p_{j(i)})\}$
is the ensemble got by superposing the $j$'th members
of each of the ensembles
with probability $p_{j(i)}$
associated with the $j$'th member of the $i$'th
ensemble.  ${\cal E}_\beta$ has density
matrix $\rho_\beta = \sum_{j(i),j(i')} \sqrt{p_{j(i)}
p_{j(i')}} |\psi_{j(i)}\rangle\langle \psi_{j(i')}|$.
In this notation, states with different $j$ cannot
interfere, but states with the same $j$ but different
$i$ can interfere.

This definition of superpositions of ensembles
allows us to
complete the identification of ensembles with vectors
by defining
$\alpha{\cal E}_\phi + \beta{\cal E}_\psi
= {\cal E}_{\alpha\phi + \beta\psi}$.  In
addition, this definition of superposition makes
a self-consistent connection between the ensemble and
superscattering pictures of time evolution, a fact
that will prove useful below.
The ensemble picture is related to
the operator sum representation of superscattering
operators described, e.g., in reference (20).

\bigskip\noindent{\bf Appendix 1}

\noindent Outline of the proof of theorem 1.
Theorem 1 follows from directly from the results of
references (1) and (3), where a detailed treatment
of high-probability subspaces may be found.  The
proof goes as follows.
If $|\psi\rangle$ is selected, with probability
$p_{|\psi\rangle}$, then
$$\sum_{|\psi\rangle}
p_{|\psi\rangle} \langle\psi| P_{\tilde{\cal H}^N}
|\psi\rangle =
{\rm tr} P_{\tilde{\cal H}^N} \rho^N$$ is just the
classical probability of the set of
high-probability sequences, and
$\rightarrow 1$ as $N\rightarrow\infty$.  As a result,
for any $\epsilon>0$, $N$ can be picked sufficiently
large so that a state picked from any stationary,
ergodic ensemble with density matrix $\rho$ has overlap
$\geq 1-\epsilon$ with some state in $\tilde{\cal H}^N$,
with
probability $\geq 1-\epsilon$.  Minimality follows
since ${\cal E}_{\tilde{\cal H}^N}$ is itself an
ensemble with density matrix $
\rho\otimes\ldots\otimes\rho$ as $N\rightarrow\infty$.
Minimality is a relatively weak property:
$\tilde{\cal H}^N$ need not be the only
minimal subspace; but all other such minimal
subspaces have approximately the same dimension as
$N\rightarrow\infty$.

\bigskip\noindent{\bf Appendix 2}

\noindent Outline of the proof of theorem 2.  There
are several ways to prove the
noisy channel theorem.  One way is to
follow along the lines
suggested in the text and analyze the channel's
effect on entangled states.  The following method of
proof is closer in spirit to the classical derivation
of channel capacity.

In the density matrix picture of the channel, the
channel has the effect,
$$|\alpha\rangle\langle\alpha| \longrightarrow \rho_\alpha
= \sum_{
i_1\ldots i_N, i'_1\ldots i'_N}
\alpha_{i_1\ldots i_N}\bar\alpha_{i'_1\ldots
i'_N} {\cal S}(|\phi_{i_1}\rangle\langle \phi_{i'_1}|)
\otimes \ldots \otimes {\cal S}(|\phi_{i_N}\rangle\langle
\phi_{i'_N}|)\quad,\eqno(2.1)$$
\noindent where the sum is taken over high-probability
sequences in which $i$ appears $\approx p_iN$ times.

Equivalently, in the ensemble picture,
$$\eqalignno{
|\alpha\rangle \longrightarrow {\cal E}_\alpha &= \{
(\sum_{i_1\ldots i_N} \alpha_{i_1\ldots i_N}
|\psi_{j_1(i_1)}\rangle \ldots |\psi_{j_N(i_N)}\rangle,
p_{j_1(i_1)}\ldots  p_{j_N(i_N)}
) \}&(2.2)\cr
&\equiv \{ (\sum_{\bf i} \alpha_{\bf i} |\psi_{\bf
j(i)}\rangle, p_{\bf j(i)} ) \}\quad,&(2.3)\cr}
$$
where the superposition ensemble is defined as
in appendix 0 and has density matrix $\rho_\alpha$.
Theorem 1 implies that as
$N\rightarrow\infty$, then with probability 1, the
states of ${\cal E}_\alpha$ are to be found in the
Hilbert space $\tilde{\cal H}_\alpha^N$ spanned by
high-probability
states of the form $$\sum_{i_1\ldots i_N}
\alpha_{i_1\ldots i_N} |\psi_{j_1(i_1)}\rangle\ldots
|\psi_{j_N(i_N)}\rangle\quad,$$
where in each term of the
superposition, $|\psi_{j(i)}\rangle$ appears $\approx
p_ip_{j(i)} N$ times.  The minimality of $\tilde{\cal
H}_\alpha$ follows as in theorem 1.
This proves the first part of theorem 2.

The dimension of the output
Hilbert space $\tilde{\cal H}_\alpha^N$ is equal to
one over the average overlap of two members of that
space:
${\rm dim}\tilde{\cal H}_\alpha^N
 = ({\rm tr_{hp}} \rho_\alpha^2)^{-1}$, where the trace
${\rm tr_{hp}}$ is taken over high-probability
sequences only.  We wish to calculate the average
dimension of the output Hilbert space over $\alpha$.
Using the fact that $<\alpha_{i_1\ldots i_N}
|\bar\alpha_{i'_1\ldots i'_N} >_\alpha = p_{i_1}\ldots
p_{i_N} \delta_{i_1i'_1}\ldots \delta_{i_Ni'_N}$,
after some algebra, we obtain
$$ <{\rm tr_{hp}} \rho_\alpha^2>_\alpha =
{\rm tr_{hp}} (\rho_{\rm out}^2)^N
+ {\rm tr_{hp}} (\rho_{\bar\alpha}^2)^N
- {\rm tr_{hp}} (\rho_{\rm i/o}^2)^N
\quad,\eqno(2.4)$$
\noindent where
$\rho_{\rm out}$ and $\rho_{\bar\alpha}$ are defined
as above,
$\rho_{\rm i/o} = \sum_i p_i{\cal S}(|\phi_i\rangle
\langle \phi_i| ) \otimes |\phi_i \rangle \langle
\phi_i|$, and $(\rho^2)^N = \rho^2\otimes\ldots\otimes
\rho^2$.  We can now use the fact
that ${\rm tr_{hp}}
(\rho^2)^N = 2^{N{\rm tr} \rho {\rm log}_2 \rho}$,
which can be simply verified
in a basis in which $\rho$ is diagonal.
We then have
$$\eqalignno{{\rm tr_{hp}} (\rho_{\rm out}^2)^N
&= 2^{N{\rm tr} \rho_{\rm out} {\rm log}_2 \rho_{\rm
out}}= 2^{-NS_{\rm out}},&(2.5)\cr
{\rm tr_{hp}} (\rho_{\bar\alpha}^2)^N
&= 2^{N{\rm tr} \rho_{\bar\alpha} {\rm log}_2
\rho_{\bar\alpha}} = 2^{-NS_{\bar\alpha}}, &(2.6)\cr
{\rm tr_{hp}} (\rho_{\rm i/o}^2)^N
&= 2^{N{\rm tr} \rho_{\rm i/o} {\rm log}_2
\rho_{\rm i/o}} = 2^{-N(\sum_i S_{\rm out}(i) + S_{\rm
in})}. &(2.7)\cr}
$$
\noindent As $N\rightarrow \infty$, $<({\rm dim}
\tilde{\cal H}^N_{\alpha})^{-1}>$ goes to the largest
of these three terms of which the third is less than
or equal to
either of the first two.  We have actually calculated
the average of the inverse of the dimension of the output
subspace: however, the standard deviation
$\sqrt{<({\rm tr}_{\rm hp} \rho_\alpha^2)^2>_\alpha -
<{\rm tr_{hp}} \rho_\alpha^2>_\alpha^2}$
is proportional to
$ (<{\rm tr_{hp}} \rho_{\rm out}^2>_\alpha
<{\rm tr_{hp}} \rho_{\bar\alpha}^2>_\alpha)^{N/2}$ and
so goes to zero
exponentially faster in $N$ than
$<{\rm tr_{hp}} \rho_{\alpha}^2>_\alpha$ except
when $S_{\bar\alpha}
= S_{\rm out}$, in which case $C_Q=0$.  As a result, the
average of the inverse is the inverse of the average, and
the average dimension of ${\rm dim}\tilde{\cal
H}^N_\alpha$ is the smaller of
$2^{-N{\rm tr} \rho_{\bar\alpha}{\rm
log}_2 \rho_{\bar\alpha}}$
and
$2^{-N{\rm tr} \rho_{\rm out}{\rm
log}_2 \rho_{\rm out}}$,
proving the second half of
theorem 2.  Note also that the standard deviation of
the dimension of $\tilde{\cal H}^N_\alpha$ as a
fraction of the average dimension also goes to zero
as $N\rightarrow\infty$, showing that almost all $\alpha$
correspond to an output space of the same
dimension.

\bigskip\noindent{\bf Appendix 3.}

\noindent Outline of the proof of theorem 3.
The high
probability subspace for this source has dimension
$2^{-N{\rm tr}\rho {\rm log}_2 \rho}$.
Encode the basis states $|\chi_i^N\rangle$ for the
source as {\it randomly chosen} orthogonal states
$|\alpha^N_i\rangle$ in the high-probability subspace of
a source that attains the channel capacity.  The channel
takes each $|\alpha^N_i\rangle$ to some state in the
ensemble ${\cal E}_{\alpha_i}$ with minimal
subspace $\tilde{\cal H}^N_{\alpha_i}$.  The average
over $\alpha_\ell$ of the
overlap $|\langle\psi_{\alpha_i} |
\psi_{\alpha_j}\rangle|$ of states
$|\psi_{\alpha_i}\rangle \in \tilde{\cal H}^N_{\alpha_i}$,
$|\psi_{\alpha_j}\rangle \in \tilde{\cal H}^N_{\alpha_j}$,
for $i\neq j$ can be calculated as in appendix 2,
and is equal to $1/{\rm dim}\tilde{\cal H}^N_{\rm out} =
2^{N{\rm tr} \rho_{\rm out} {\rm log}_2 \rho_{\rm out}}$.
If $P^N_{\alpha_i}$ is the projection operator onto
$\tilde{\cal H}^N_{\alpha_i}$, we have
$${\rm tr} P^N_{\alpha_i} P^N_{\alpha_j} = 2^{-N(-
{\rm tr} \rho_{\rm out} {\rm log}_2 \rho_{\rm out}+
{\rm tr} \rho_{\bar\alpha} {\rm log}_2 \rho_{\bar\alpha})}
= 2^{-NC_Q}\quad.\eqno(3.1)$$
\noindent That is, as $N\rightarrow\infty$, the overlap
between any two individual output subspaces $\rightarrow 0$
as long as the quantum channel capacity is not zero.
The dimension of the direct sum of the output
subspaces remains less than or equal to the dimension of
${\cal H}^N_{\rm out}$ if and only if $-{\rm tr}
\rho {\rm log}_2 \rho \leq C_Q$:
$$\eqalignno{{\rm dim}~\oplus \sum_i \tilde{\cal
H}^N_{\alpha_i} &\rightarrow 2^{-N({\rm tr} \rho {\rm
log}_2 \rho + \rho_{\bar\alpha} {\rm log}_2
\rho_{\bar\alpha})}\cr
&= 2^  {N(-{\rm tr} \rho_{\rm out} {\rm
log}_2 \rho_{\rm out} - \zeta)}\quad,&(3.2)\cr}$$
\noindent where $\zeta = C_Q -(-{\rm tr}\rho {\rm log}_2
\rho)$.  So if $\zeta\geq 0$, the source entropy does not
exceed the channel capacity, and the output states
corresponding to different input basis states all fall
in distinct subspaces.  The overlap of any one output
subspace with the direct sum of all the remaining subspaces
goes as $2^{-N\zeta}$.
If $\zeta<0$, the output
subspaces overlap and no unique decoding is possible.
This proves that $C_Q$ is an upper limit on the channel
capacity for `typical' codewords belonging to the
high-probability subspace
(i.e., for a set of measure
1 as $N\rightarrow\infty$), but it does not rule
out the possibility of the use of a set of codewords of
measure 0.

In the case $\zeta\geq 0$,
a unitary decoding transformation can now be applied to the
output states to put each vector $|\psi^N_{\alpha_i}\rangle
 \in \tilde{\cal H}^N_{\alpha_i}$ into the form
$|\chi_i^N\rangle\otimes|\psi^N\rangle$, in which
vectors in different output subspaces but with the same
$|\psi_{\bf j(i)}\rangle$ in (2.3) give the same
$|\psi^N\rangle$.  Because of the
asymptotic orthogonality of the output spaces, this
decoding recreates
$|\chi^N_i\rangle$ with fidelity arbitrarily close to 1 as
$N\rightarrow\infty$.  The crucial point is that this
decoding also recreates {\it superpositions} of input states
with fidelity $\rightarrow 1$ as $ N\rightarrow\infty$:
by going to the
ensemble picture, it can be verified that $\sum_k
\gamma_k |\chi_k^N\rangle$ is mapped to an ensemble
$\{(\sum_k
\gamma_k|\chi_k^N\rangle\otimes |\psi^N\rangle,
p_{|\psi\rangle})\}$.   The steps are as follows.
First, encoding:
$$\sum_k\gamma_k|\chi^N_k\rangle
\longrightarrow
\sum_k\gamma_k \sum_{i_1\ldots i_N} \alpha^k_{i_1\ldots
i_N} |\phi_{i_1}\rangle \ldots |\phi_{i_N}\rangle
\eqno(3.3a)$$
\noindent Next, the effect of the channel:
$$\longrightarrow\{~(
\sum_k\gamma_k \sum_{i_1\ldots i_N} \alpha^k_{i_1\ldots
i_N} |\psi_{j_1(i_1)}\rangle \ldots |\psi_{j_N(i_N)}\rangle
{}~,~
 p_{j_1(i_1)}\ldots  p_{j_N(i_N)})
{}~\} \eqno(3.3b)$$
\noindent Finally, decoding:
$$\eqalignno{&\longrightarrow\{~(
\sum_k\gamma_k |\chi_k\rangle\otimes
\sum_{i_1\ldots i_N} \beta_{i_1\ldots
i_N} |\psi_{j_1(i_1)}\rangle \ldots |\psi_{j_N(i_N)}\rangle
{}~,~
 p_{j_1(i_1)}\ldots  p_{j_N(i_N)})
{}~\}\cr
&= \{ (\sum_k\gamma_k|\chi_k\rangle\otimes|\psi^N\rangle
{}~,~ p_\psi)\}
\quad.&(3.3d)\cr}$$

\noindent The fact that
the decoding process faithfully recreates
superpositions can also be verified in the density
matrix picture by using the correspondence in appendix
2.  Since the encoding and decoding preserves pure
states with their phases, it also preserves mixed
states and any entanglement between the input
state and another quantum system.

This proves the if part of the theorem.  The only if
part for codewords from the high-probability subspace
was proved above.  This proves the theorem as
stated.

The limits set by theorems 2 and 3 hold only for
codewords from the high-probability set: by using
codewords taken from the set of measure zero, it
may be possible to improve on these limits.$^{21}$
A simple example of how this may be done is given by
the method of theorem 3 itself: block together the
quantum symbols (e.g., qubits) in groups of $\ell$,
and regard each group of $\ell$ as a new, composite
symbol.  The minimization procedure used for finding
the quantum channel capacity in general yields a
different, potentially higher channel capacity for
codes composed of the composite symbols.

\vfill\eject
$ $
\vskip 1in
\noindent{\bf Figure 1}

\vskip .4in
\centerline{{\bf Noise}$\quad$  {\bf Decoherence}$\quad$}

\centerline{$\downarrow$\hskip1cm$\uparrow$\hskip1cm}

\centerline{$|\psi\rangle \longrightarrow$ {\bf Encoder}
$\longrightarrow C(|\psi\rangle)
\longrightarrow$
{\bf Channel} $\rightarrow N\big(C(|\psi\rangle)\big) \rightarrow$
{\bf Decoder}
$\rightarrow |\psi\rangle +$ {\bf Noise}}

\vskip 1cm
\noindent {\it Figure 1: Diagram of the noisy, decoherent
quantum channel.}
To send an arbitrary quantum state $|\psi\rangle$ down the channel,
first encode it in a redundant form $C(|\psi\rangle)$.  The encoded
state is sent down the channel, where it is
subjected to noise and decoherence.  The arrows indicate that noise
is added to the signal, while decoherence arises from the
environment getting information about the signal.
The noisy, decoherent signal $N\big(C(|\psi\rangle)\big)$
is then fed through a decoder that
recreates the original state together with extra random information
that depends on what errors occurred.
\vfill\eject\end